\documentclass[%
reprint, twocoloumn,
amsmath,amssymb,article,showpacs,  
aps,
]{revtex4-2}
\usepackage{graphicx}
\usepackage{dcolumn}
\usepackage{blindtext}
\usepackage{bm}
\usepackage{tikz}
\usepackage{pgfplots}
\pgfplotsset{compat=1.14}
\usepackage{adjustbox}
\usetikzlibrary{decorations.pathreplacing}
\usepackage{amssymb,float}
\usepackage{float}
\usepackage{amsmath}
\usepackage{lipsum}
\usepackage{rotating}
\usepackage{color}
\usepackage{xcolor}
\usepackage{pgfplots}

\usepackage{epsfig}
\usepackage{epstopdf}
\usepackage{bm}
\usepackage{textcomp}
\usepackage{array}
\usepackage{upgreek}
\usepackage{caption}
\usepackage{ctable}
\usepackage{xfrac}
\usepackage{mathtools}
\usepackage{amsfonts}
\usepackage{subfigure}
\usepackage{hyperref}
\usepackage{booktabs}
\usepackage{multirow}
\usepackage{tikz}        
\usepackage{lineno}      
\usepackage{cleveref}    

\hypersetup{
        colorlinks=true,
        linkcolor=red,
        filecolor=magenta,
        urlcolor=blue,
        pdftitle={}
        }
\DeclareRobustCommand{\orcidicon}{%
	\begin{tikzpicture}
	\draw[lime, fill=lime] (0,0) 
	circle [radius=0.16] 
	node[white] {{\fontfamily{qag}\selectfont \tiny ID}};
	\draw[white, fill=white] (-0.0625,0.006) 
	circle [radius=0.007];
	\end{tikzpicture}
	\hspace{-2.9mm}
}

\foreach \x in {A, ..., Z}{\expandafter\xdef\csname orcid\x\endcsname{\noexpand\href{https://orcid.org/\csname orcidauthor\x\endcsname}
		{\noexpand\orcidicon}}
}

\DeclareRobustCommand{\orcidicon}{%
	\begin{tikzpicture}
	\draw[lime, fill=lime] (0,0) 
	circle [radius=0.15] 
	node[white] {{\fontfamily{qag}\selectfont \tiny ID}};
	\draw[white, fill=white] (-0.0625,0.005) 
	circle [radius=0.007];
	\end{tikzpicture}
	\hspace{-2.9mm}
}

\foreach \x in {A, ..., Z}{\expandafter\xdef\csname orcid\x\endcsname{\noexpand\href{https://orcid.org/\csname orcidauthor\x\endcsname}
		{\noexpand\orcidicon}}
}

\usepackage{etoolbox}
\apptocmd{\sloppy}{\hbadness 10000\relax}{}{}

\usepackage{silence}
\usepackage{hyperref}
\usetikzlibrary{animations}
\begin{document}
\title{Mg$_{2}$Si and Ca$_{2}$Si semiconductors for photovoltaic applications: Calculations based on density-functional theory and the Bethe-Salpeter equation}
	\author{Vinod Kumar Solet \orcidA{}}
	\email{vsolet5@gmail.com}
	\author{Sudhir K. Pandey \orcidB{}}
	\email{sudhir@iitmandi.ac.in}
	\affiliation{School of Mechanical and Materials Engineering, Indian Institute of Technology Mandi, Kamand $-$ 175075, India}
	\date{\today}
	
\begin{abstract}
We conduct a comprehensive assessment of the electronic and optical properties, as well as photovoltaic ({\small{PV}}) performance parameters, for low-cost, nontoxic Mg$_{2}$Si and Ca$_{2}$Si using density-functional theory (DFT) and Bethe-Salpeter equation (BSE) based methods. The band-gap for Mg$_{2}$Si (Ca$_{2}$Si) is found to be in the range of 0.25-0.6 (0.57-0.96) eV when PBE, PBEsol and mBJ functionals are used. Effective masses at last-filled valence and first-empty conduction bands are in the range of 0.14-0.17 (1.17-1.25)m$_{e}$ and 0.27-0.29 (0.3-0.41)m$_{e}$, respectively. In the \textit{independent-particle} approximation (IPA), the real and imaginary parts of dielectric function show maximum values of $\sim$50 (16.3) at $\sim$2.6 (1.0) eV and $\sim$61 (16.2) at $\sim$3.24 (3.4) eV, respectively. Within BSE, these respective values change to $\sim$59 (17) at $\sim$2.5 (0.86) eV and $\sim$65 (16.6) at $\sim$2.68 (3.1) eV. The excitonic effect is found to be crucial in understanding the experimental optical spectra of Mg$_{2}$Si. However, this effect is relatively weaker in Ca$_{2}$Si. In solar spectrum active region, the highest absorption coefficient and lowest reflectivity change from $\sim$1.5 (0.75)$\times$$10^{6}$ $cm^{-1}$ and $\sim$0.39 (0.3) in IPA, respectively, to $\sim$1.6 (0.88)$\times$$10^{6}$ $cm^{-1}$ and $\sim$0.41(0.27) in BSE. Present study highlights the importance of different levels of theoretical approximations for obtaining the optical spectroscopy data of silicides with a high level of accuracy. Finally, we have evaluated {\small{PV}} efficiency by using spectroscopic limited maximum efficiency ({\small{SLME}}) calculation. On the top of radiative recombination, we have also incorporated non-radiative carrier recombination at a defect trap state via Shockley-Read-Hall ({\small{SRH}}) mechanism to evaluate the efficiency. Among the studied defects, the interstitial Mg (Si) is identified as the most stable in Mg$_{2}$Si  (Ca$_{2}$Si), and this provides {\small{SRH}} lifetime of $\sim$2 $\mu s$ (11.3 $ms$). The estimated maximum {\small{SLME}} using BSE absorption spectrum is $\sim$1.3 (31.2)\%, which decreases to $\sim$1.2 (28.5)\% due to {\small{SRH}} recombination. The present study suggests that Ca$_{2}$Si (Mg$_{2}$Si) is a potential candidate for single-junction (bottom cell in multi-junction) thin-film PV devices.
    
\end{abstract}

\maketitle
\section{Introduction} 
\setlength{\parindent}{3em}

Semiconductors are playing a pivotal role in the current technological revolution. The spectroscopy of light-matter interaction in semiconductors is an intriguing topic in both modern research and industrial applications \cite{rondinelli2015predicting}. One of the most significant technologies for generating renewable energy is {\small{PV}} solar cells. Essentially, a common question when designing solar cells is the amount of incident sunlight that the optically active materials can absorb, as well as how to adjust these materials to absorb more light and generate a large number of electron-hole ($e$-$h$) pairs. It also becomes critical to understand how to optimize materials that efficiently convert these $e$-$h$ pairs into useful electricity. The response function calculation \cite{cohen1988electronic} is a fundamental study that particularly provides a theoretical spectroscopic analysis of {\small{PV}} semiconducting materials.

The past few decades have seen the thorough use of $first$ $principles$ based DFT and Fermi's golden rule to investigate the optical absorption of many semiconductors \cite{levine1989linear,onida2002electronic}. Unfortunately, DFT calculations do not describe the key aspect of optical absorption, which is $e$-$h$ interaction, because they only deal with single-particle states \cite{bechstedt2016many}. An absorbing photon excites a valence electron into the conduction band, leaving a hole behind in the valence band; thus, $e$-$h$ pair is created. Single-particle theories treat this created $e$-$h$ pair as independent, non-interacting particles. These theory-based optical spectra, estimated within \textit{independent-particle} approximation (IPA), often have significant deviations from experiment \cite{rohlfing2000electron,begum2021theoretical,marinopoulos2011local}. However, for an interacting system, it has been assumed that the bound states of an excited $quasielectron$ and a $quasihole$ (excitons) are the missing ingredients for a significant comparison between the theoretical and experimentally measured optical spectra \cite{benedict1998optical,chang2000excitons,arnaud2001local,onida2002electronic}. To obtain an accurate optical response, one must consider the \textit{screened} Coulomb interaction in the excited $e$-$h$ pairs. In particular, the Bethe-Salpeter equation (BSE) \cite{salpeter1951relativistic}, an effective two $quasiparticle$ equation for polarizability \cite{albrecht1998ab,onida2002electronic}, can be solved using many-body perturbation theory (MBPT) based on Green’s function approach in an \textit{ab initio} framework \cite{rohlfing2000electron,vorwerk2019bethe}. 

Efficient solar devices do not always rely on optical parameters \cite{yu2012identification}; they also depend on the type of band-gap as well as various solar cell parameters \cite{dalal1977design,yu2012identification,kirchartz2018makes}. Some successful theoretical studies using the $first$ $principles$ approach have identified new solar materials with a high power conversion efficiency \cite{lee2014computational,yin2014unique}. The Shockley-Queisser (SQ) model \cite{shockley2004detailed} is used widely to predict the theoretical efficiency of single-junctional solar cells. In this approach, the efficiency is directly related to the band-gap of the light-absorbing layer, while other properties such as optical absorption, carrier transport, and lifetimes are assumed to be ideal. The SQ-model can be further improved by considering optical absorption in {\small{SLME}} model \cite{yu2012identification}. Generally, any excited or injected charge carrier that does not leave the semiconducting material can undergo various types of recombination \cite{nelson2003the,smets2016solar}. The difference between SQ limit or {\small{SLME}} and the efficiencies of actual solar devices arise mainly from additional irreversible processes, such as non-radiative $e$-$h$ recombination. Therefore, we have also calculated the efficiency by including losses of $e$-$h$ pairs inside solar absorbers both from radiative and non-radiative processes \cite{kirchartz2018makes,kim2020upper}. The current density due to radiative recombination can be easily estimated, but one needs to know about defect properties in semiconductors to determine non-radiative current density \cite{shi2012ab, xu2021defect}. In {\small{PV}} solar devices, the trap states or recombination centers of a defect in the band-gap help photo-excited carriers to recombine non-radiatively, also commonly referred to as {\small{SRH}} recombination \cite{shockley1952statistics,hall1952electron,smets2016solar}. In our calculation, the {\small{SRH}} carrier lifetime limits the efficiency based on the electronic dispersion, optical absorption, and thermodynamical properties of defects of an absorbing material.  

This paper chooses solar materials based on their environmental friendliness, lower cost, availability on earth, and non-toxic characteristics. Silicon-based materials can be one of the best choices used in solar cells. In this category, researchers gain a lot of interest in semiconducting silicides \cite{d2000silicides,borisenko2013semiconducting} in thin-film solar technology due to their high optical absorbance \cite{liu2006thin,makita2006beta}. For instance, the Mg$_{2}$Si semiconductor \cite{borisenko2013semiconducting} is one of the silicides that has been studied the most. It has experimental indirect band gaps in the range of 0.61 to 0.8 eV \cite{udono2015crystal,scouler1969optical,morris1958semiconducting,kato2011optoelectronic,mahan1996semiconducting,tamura2007melt,heller1962seebeck} and a large absorption coefficient from reflectance measurements of $\sim$3$\times$10$^{5}$ $cm^{-1}$ at 2.5 eV of visible energy \cite{scouler1969optical}. However, in theory, the indirect band gaps for Mg$_{2}$Si lie in the range of $\sim$0.118-0.53 eV \cite{corkill1993structural,imai2003electronic,aymerich1970pseudopotential,au1969electronic}. Various other experiments such as spectroscopic ellipsometry \cite{kato2011optoelectronic} and reflectance \cite{sobolev1972reflectivity,mcwilliams1963infrared,vazquez1968electroreflectance} measurements were performed to obtain optical properties for Mg$_{2}$Si. Another same class of silicide is cubic Ca$_{2}$Si, which is found to have theoretical direct band gaps of 0.56 \cite{migas2003comparative} and 1.16 \cite{lebegue2005calculated} eV from DFT and \textit{GW} approximation, respectively. However, the full investigation of both silicides for solar applications has yet to be done, and the appropriate band gaps provide a promising outlook for future research into their solar application properties. But the available theoretical gaps are not consistent with each other. So, it makes sense to estimate a band gap for Mg$_{2}$Si that is similar to what has been observed in experiments. The band gap calculation in DFT depends a lot on the exchange and correlation ({\small{XC}}) potentials. One thing also to keep in mind is that we work here on the theoretical (electronic and optical properties) as well as application aspects for these two silicides. Therefore, we first validate the different approximations based on IPA and MBPT for theoretical aspect in Mg$_{2}$Si using the available experimental data. Based on this validation, we explore its practical applications in {\small{PV}} technology. We then apply the same methodology to Ca$_{2}$Si, despite the lack of experimental data, to provide insights into its potential applications.

Therefore, this study includes a $first$ $principles$ analysis using two functionals (PBE, PBEsol) and mBJ potential to estimate the electronic and optical properties for Mg$_{2}$Si and Ca$_{2}$Si, with the aim of optimizing their solar efficiency. The predicted fundamental band-gaps range for Mg$_{2}$Si and Ca$_{2}$Si are $\sim$0.25-0.60 eV and $\sim$0.57-0.96 eV, respectively. The evaluated optical gap from mBJ is $\sim$2.45 and 0.96 eV, respectively. We also evaluate the density of states effective masses of charge carriers at high-symmetric points. The optical properties are calculated without and with considering $e$-$h$ interaction in IPA (PBE, mBJ and PBE+scissor) and BSE methods, respectively. All optical parameters show broad spectra in the visible light. However higher absorption can’t guarantee an efficient single-junction solar cell. Therefore, we estimate the efficiency by incorporating various $e$-$h$ recombination processes. At 300 K, the highest predicted efficiency using BSE absorption coefficient by explicitly considering only radiative recombination ({\small{SLME}})  is $\sim$31.2\% for Ca$_{2}$Si and $\sim$1.3\% for Mg$_{2}$Si. The point-defect calculation mainly yields the {\small{SRH}} lifetime required to calculate the non-radiative recombination. Among the defects analyzed, the interstitial Mg (Si) defect is found to be most stable in Mg$_{2}$Si (Ca$_{2}$Si). Accounting for thermodynamic properties of defect, the {\small{SRH}} lifetime is estimated to be $\sim$2 $\mu s$ (11.3 $ms$). Consequently, the maximum possible trap-limited conversion ({\small{TLC}}) efficiency at 300 K using BSE absorption spectra is predicted to be $\sim$1.2 (28.5)\%. As a result, Ca$_{2}$Si can be used to make thin-film single-junction solar cells. Moreover, narrow-gap Mg$_{2}$Si, with its high absorption is a suitable {\small{PV}} material for fabricating a bottom layer in multi-junction cell.

\section{Computational details}
 
Electronic structure calculation was performed by using of augmented plane wave plus local orbitals ({\small{APW+lo}}) method within {\small{DFT}} as implemented in {\small\texttt{WIEN2k}} \cite{blaha2020wien2k} code. The muffin-tin sphere radii (R$_{MT}$) value was kept fixed to 2.5 Bohr for all atoms. Also, a 10$\times$10$\times$10 \textbf{k}-mesh size and convergance criteria of $10^{-4}$ Ry/cell for total energy had to be considered for solving the Kohn-Sham (KS) equation. Further two different functionals,\textit{ i.e.,} PBE \cite{perdew1996generalized} and PBEsol \cite{perdew2008restoring} were considered as a {\small{XC}} part of the KS equation. The inherent band-gap underestimation of DFT was corrected by the using of Tran-Blaha modified Becke-Johnson (TB-mBJ) potential \cite{tran2009accurate}. 

The optical dielectric function, at IPA level, was calculated using the {\small\texttt{WIEN2k}} code. The $independent$-$particle$ polarizability, which uses KS orbitals and eigenvalues, required for the calculation of dielectric function was estimated within the random-phase approximation (RPA), which is also known as the KS-RPA. Meanwhile, the \textbf{k}-mesh was built into a large 5000 \textbf{k}-points over the full Brillouin zone to obtain the momentum matrix elements. The Lorentzian broadening value of 50 meV was used. The components of an effective mass tensor using the degenerate perturbation approach were obtained by the {\small\texttt{mstar}} code implemented in {\small\texttt{WIEN2k}} software \cite{rubel2021perturbation}. Effective mass calculation in this framework is very sensitive to the momentum matrix elements that contain upper energy bands. Thus, these matrix elements were obtained within the maximum energy of 10 Ry, which corresponds to the 236 and 164 bands for Ca$_{2}$Si and Mg$_{2}$Si, respectively. The tolerance parameter of degeneracy energy was used as $10^{-5}$ Ha.

Optical properties including $e$-$h$ interactions using MBPT were estimated from the all-electron full-potential {\texttt{exciting}} code through the solution of BSE within the Tamm-Dancoff approximation (TDA) \cite{vorwerk2019bethe}. Initially, the ground state eigenenergies and eigenstates from PBE functional were obtained at a \textbf{k} grid of $12\times 12\times 12$. A $5\times 5\times 5$ \textbf{k}-mesh size shifted by (0.097, 0.273, 0.493) was employed for the BSE calculation. For screening calculation, a $5\times 5\times 5$ \textbf{q}-mesh size was used. We used 50 empty states for both silicides to construct the excitonic Hamiltonian. The energy threshold for including the local field effects in the excited states, $\mid$\textbf{G} + \textbf{q}$\mid$$_{max}$, was taken into account upto a cutoff of 3.0 $a_{0}^{-1}$. A Lorentzian lifetime broadening of 50 meV was applied to the optical spectra to obtain comparable results with the experiment. For both compounds, 100 empty states were adopted in the RPA calculation for the $screened$ Coulomb interaction between electron and hole. A scissor shift of 0.55 and 0.4 eV was applied to the optical spectra in order to mimic the $quasiparticle$ (QP) correction in  Mg$_{2}$Si and Ca$_{2}$Si, respectively. Finally, the considered transition space includes all occupied and the first 20 unoccupied bands in order to diagonalize the BSE Hamiltonian for both compounds.

The calculation of point defects were performed using DFT within $2\times 2\times 2$ size of supercell containing of total 96 atoms in pristine structure for both compounds using the {\small\texttt{WIEN2k}} code \cite{blaha2020wien2k}. The defects were simulated by removing and/or adding the constituent atoms from and/or to the pristine structure. The PBE functional was used \cite{perdew1996generalized}, and BZ was sampled with a \textbf{k}-mesh size of $4\times 4\times 4$. The self-consistency calculations continued until the structure was fully relaxed through structural optimization.

\section{Results and discussion}
\subsection{\label{sec:level2}Electronic properties}

Here, the electronic and optical properties of Mg$_{2}$Si and Ca$_{2}$Si compounds have been investigated in their anti-fluorite structure with a space group of Fm$\bar{3}$m (No. 225). The used value of lattice parameters in the entire calculations for Mg$_{2}$Si and Ca$_{2}$Si is 6.35 \AA, and 7.16 \AA, respectively. The value of lattice parameter for Mg$_{2}$Si is very close to the experimental value \cite{pulikkotil2012doping}. The primitive unit cell of both compounds consists of three atoms, in which one Si atom is located at the (0, 0, 0) and two Mg (Ca) atoms are situated at the position of (0.25, 0.25, 0.25) and (0.75, 0.75, 0.75). All these above structural parameters are directly taken from the {\small\texttt{AFLOW}} database \cite{aurtarolo2012aflow}.

The calculation of the band gap from electronic dispersion offers a promising method for finding suitable {\small{PV}} materials. Optical processes have been related to vertical (or direct) transitions and number of photo-excited electrons in conduction bands strongly depends on the band gap. The optical dielectric function in Eq.~\hyperref[eq:ipa]{\eqref{eq:ipa}} is also directly related to the eigenvalues of occupied and unoccupied states. Thus both direct and indirect gaps are critical for {\small{PV}} device performance, making it essential to provide an accurate predictions for gaps. The DFT band structure gives accurate information regarding the shape of energy bands, but a major problem with this theory is that it generally underestimates the band-gap of semiconductors or insulators. This is due to its improper treatment of the {\small{XC}} potential, which includes factors such as the self-interaction error in GGA or LDA and the absence of density-functional derivative discontinuity in the {\small{XC}} energy \cite{perdew1983physical}. In order to properly handle the exchange potential, the mBJ potential \cite{tran2009accurate} is used to obtain the best possible band gaps for both silicides, as this approach is known to yield more precise gaps commonly observed in many semiconductors \cite{kim2010towards,koller2011merits}.

\begin{figure}[h]
\includegraphics[width=8.7cm, height=5.8cm]{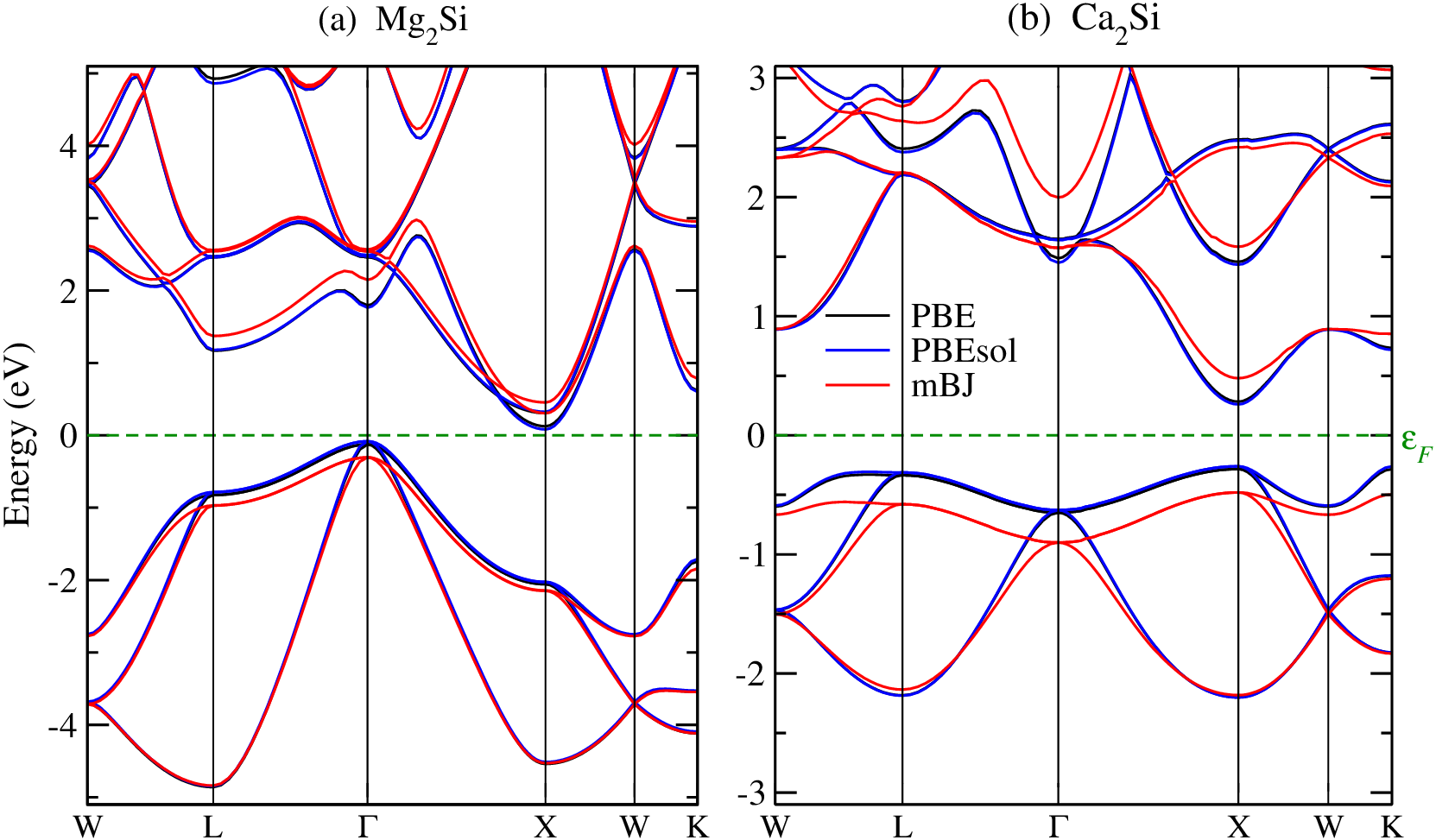} 
\caption{Electronic band dispersion for (a) Mg$_{2}$Si and (b) Ca$_{2}$Si from PBE and PBEsol functionals, and mBJ potential.} 
\label{fig:disp}
\end{figure}

Therefore, the electronic dispersions for both silicides have been calculated from PBE \& PBEsol functional, and mBJ potential, which are presented in Fig.~\ref{fig:disp}. Table~\ref{table:bandgap} summarizes the fundamental and optical band gaps obtained from the different functionals, along with the available experimental gaps for Mg$_{2}$Si. Unfortunately, experimental band gap is not available for cubic Ca$_{2}$Si. It is noticed that the band gap increases in the silicides by replacing the light Mg atom with a heavy Ca atom. In Fig.~\hyperref[fig:disp]{\ref{fig:disp}(a)}, Mg$_{2}$Si is found to be an indirect band gap semiconductor, where highest energy level of valence band (VB), $\varepsilon_{v}$, and lowest energy level of conduction band (CB), $\varepsilon_{c}$, are at $\Gamma$- and X-point, respectively. With PBE (PBEsol) the indirect gap is $\sim$0.25 (0.16) eV, 59-69 (74-80)\% smaller than the experimental gaps of $\sim$0.61-0.8 eV \cite{udono2015crystal,scouler1969optical,morris1958semiconducting,kato2011optoelectronic,mahan1996semiconducting,tamura2007melt,heller1962seebeck}. In contrast, an indirect gap of $\sim$0.6 eV within mBJ aligns well with the experimental gaps. The estimated value of an optical gap at $\Gamma$-point from PBE (PBEsol) is $\sim$1.92 (1.76) eV, while it is $\sim$2.45 eV from mBJ. Reflectivity experiment at 77 K \cite{scouler1969optical} and electro-reflectance measurement at room temperature \cite{vazquez1968electroreflectance} revealed an optical gap of $\sim$2.17 and 2.27 eV, respectively. Next, from Fig.~\hyperref[fig:disp]{\ref{fig:disp}(b)}, Ca$_{2}$Si is a direct band gap semiconductor with a value of $\sim$0.57 (0.52) eV at a $X$-point obtained from the PBE (PBEsol) functional, and the calculated gap from the mBJ potential is $\sim$0.96 eV. The VB near Fermi energy ($\varepsilon_{F}$) seems to be less (more) dispersive in the direction of $\Gamma$-$L$ ($\Gamma$-$X$) and $X$-$\Gamma$ ($X$-$W$) for Mg$_{2}$Si and Ca$_{2}$Si, respectively. These behaviors of bands in the respective directions are crucial for understanding the optical properties of both studied silicides. One can also notice in both materials that the PBE and PBEsol provide nearly comparable band structures. Therefore, further properties that depend on dispersion can be explained by taking only one of the functionals, and we consider PBE.  

\begin{table}
\caption{Band gaps in eV from PBE, PBEsol and mBJ. The fundamental and optical (in brackets) gaps for Mg$_{2}$Si are compared to experiment values. For Ca$_{2}$Si, both fundamental and optical gaps are the same.}
\resizebox{0.49\textwidth}{!}{%
\setlength{\tabcolsep}{6pt}
\begin{tabular}{@{\extracolsep{\fill}}c c c |c c c c c c c c c} 
\hline\hline
\multicolumn{1}{c}{Methods} & & \multicolumn{0}{c}{Mg$_{2}$Si} &  \multicolumn{1}{c}{Ca$_{2}$Si} \\  
 && \multicolumn{1}{c}{$\Gamma$-$X$ ($\Gamma$-$\Gamma$)} & $X$-$X$ \\
 \hline
PBE functional && 0.25 (1.92) &  0.57  \\
PBEsol functional && 0.16 (1.76) &  0.52  \\
mBJ potential && 0.60 (2.45) &  0.96  \\
Experimental  && 0.61-0.8 \cite{udono2015crystal,scouler1969optical,morris1958semiconducting,kato2011optoelectronic,mahan1996semiconducting,tamura2007melt,heller1962seebeck} (2.17-2.27 \cite{scouler1969optical,vazquez1968electroreflectance}) &  Not available \\
                            
\hline\hline 
\end{tabular}}
\label{table:bandgap}
\end{table}

\begin{figure}
\includegraphics[width=8.7cm, height=7cm]{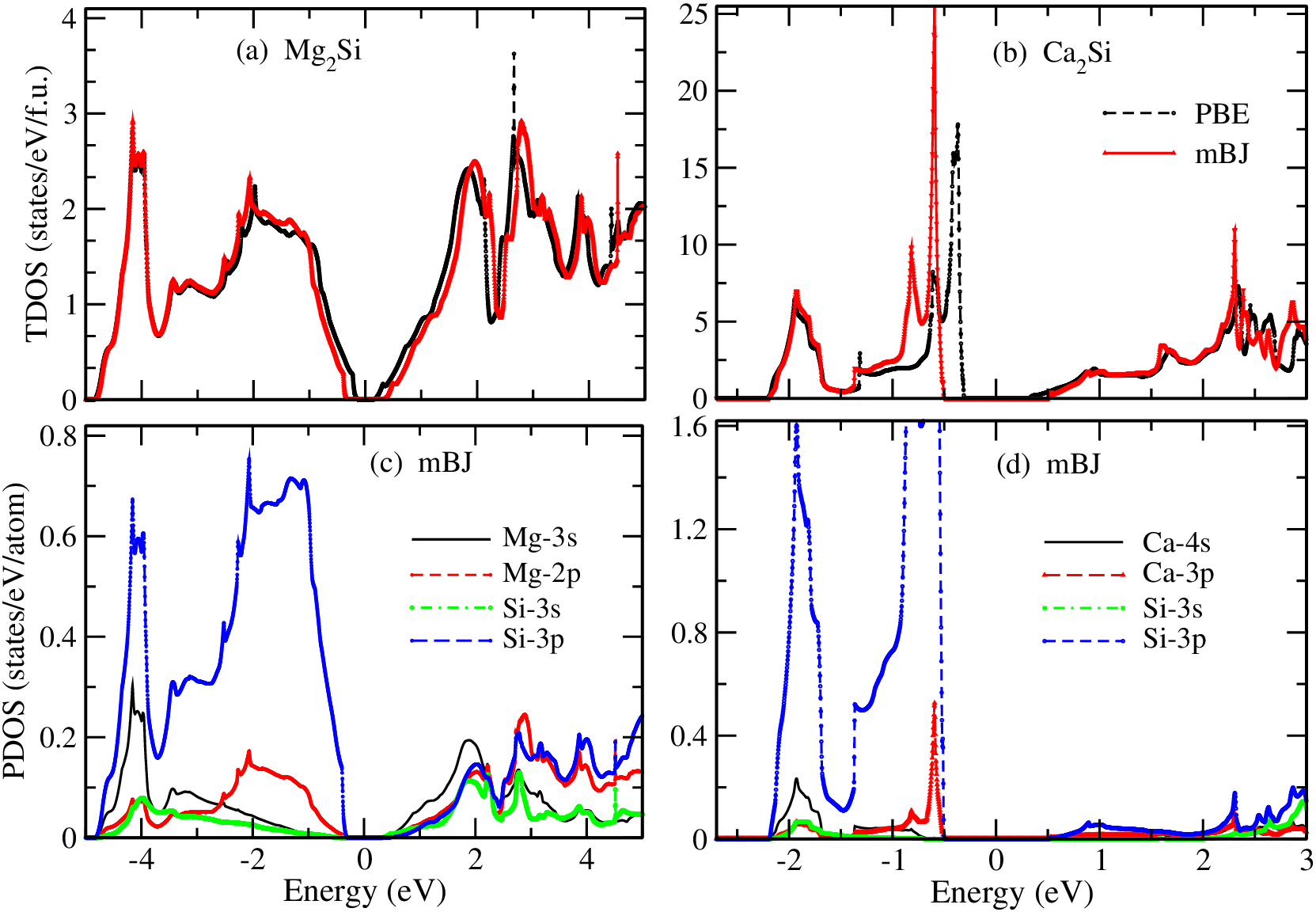} 
\caption{{Total/partial density of states (T/PDOS) plots for Mg$_{2}$Si and Ca$_{2}$Si.}} 
\label{fig:dos}
\end{figure}
 
The transition of electrons from filled to empty bands is critical to the interaction of light with semiconductors. As in Eq.~\hyperref[eq:ipa]{\eqref{eq:ipa}}, the success of such transitions depends on the number of available states as well as the transition probability of electrons from occupied to unoccupied states at a particular photon energy. Thus, in Fig.~\ref{fig:dos}, the total density of states (TDOS) along with the partial density of states (PDOS) plots have been shown. Both compounds show a band gap near the $\varepsilon_{F}$, which is a nature of being semiconducting compounds. For Mg$_{2}$Si, one can notice the sharp rise in states at the topmost (bottom-most) region of the VB (CB), which is a good sign for solar materials. Except peak intensities, there is a slightly change in PBE states with the inclusion of mBJ potential. Two maximum peaks in the VB region are located around the energies of $\sim$$-$1.95 and $\sim$$-$4.17 eV. In the CB region, a largest peak is found around the energy of $\sim$2.7 eV. Similarly, in Fig.~\hyperref[fig:dos]{\ref{fig:dos}(b)} for Ca$_{2}$Si, there have been sharply increasing states in the vicinity of the $\varepsilon_{F}$ on the VB side. These VB states are shifted by $\sim$0.2 eV on the lower energy side by adding mBJ in PBE, while states in CB have not changed significantly. The maximal peaks in the VB and CB regions from PBE (mBJ) are situated at nearly $-$0.35 (0.6) eV and $\sim$2.3 eV, respectively. The PDOS obtained from mBJ has been shown, in which the electronic states of VB are significantly contributed by the Si-3p orbital. In Fig.~\hyperref[fig:dos]{\ref{fig:dos}(c)}, the gap is mainly coming from the Si-3p and Si-3s orbitals, respectively, in the VB and CB regions. Similarly, in Fig.~\hyperref[fig:dos]{\ref{fig:dos}(d)}, the whole VB and CB regions are rich in Si-3p character with a low contribution from Ca-4s,3p states. The Si-3p and Ca-3p states in the VB and CB regions, respectively, have been responsible for creating the band gap and may also be actively involved in the photo-conversion process. 

The effective mass of the charge carriers (electrons and holes) in the vicinities of $\varepsilon_{c}$ and $\varepsilon_{v}$ is of great importance for analyzing the carrier transport such as mobilities and the current density in the {\small{PV}} materials \cite{ashcroft1976,rubel2021perturbation}. The density of states effective mass (m$_{dos}^{*}$) of the charge carriers directly affect the electrical conductivity \cite{spitzer1957determination}, and depletion width of the $pn$-junction solar cells based on intrinsic concentration \cite{ashcroft1976,kasap2006principles,green1990intrinsic}. Therefore, in Table~\ref{table:effmass}, the m$_{dos}^{*}$ is calculated at high-symmetric points for the last filled VB and first empty CB in the vicinity of $\varepsilon_{F}$. The m$_{dos}^{*}$ is obtained from the geometric mean of the principal components of the effective mass tensor ($m_{1}^{*}$, $m_{2}^{*}$, $m_{3}^{*}$), \textit{i.e.,} $\sqrt[3]{m_{1}^{*}\times m_{2}^{*}\times m_{3}^{*}}$ \cite{rubel2021perturbation}. The m$_{dos}^{*}$ at $\varepsilon_{v}$ is observed to be $\sim$0.14-0.17m$_{e}$ and 1.17-1.25m$_{e}$, whereas in $\varepsilon_{c}$, it is $\sim$0.27-0.29m$_{e}$ and $\sim$0.30-0.41m$_{e}$ for Mg$_{2}$Si and Ca$_{2}$Si, respectively. The estimated m$_{dos}^{*}$ of electron for Mg$_{2}$Si at $\varepsilon_{c}$ is well matched with previously available theoretical effective mass (m$_{\perp}^{*}$) of 0.25m$_{e}$ \cite{lee1964electronic}. The negative (positive) value of m$_{dos}^{*}$ at VB (CB) corresponds to a hole (electron) effective mass \cite{schafer2002semiconductor}. The experimental effective masses obtained by \textit{Morris et al.} \cite{morris1958semiconducting} are 0.47m$_{e}$ (for electron) and 0.87m$_{e}$ (for hole), while those by \textit{Heller et al.} \cite{heller1962seebeck} are 0.5m$_{e}$ (for electron) and 2.0m$_{e}$ (for hole). In another study, \textit{Udono et al.} \cite{udono2015crystal} found effective masses of 0.57m$_{e}$ (for electron) and 1.07m$_{e}$ (for hole), while \textit{Nolas et al.} \cite{nolas2007transport} found 0.51m$_{e}$ (for electron). In Table~\ref{table:effmass}, the m$_{dos}^{*}$ of $p$-type carriers in VB is larger than that of $n$-type carriers (similar trend as observed in experiment), and this is because the slop of DOS below $\varepsilon_{F}$ exceeds that above. One possible reason for the difference between the calculated and experimental m$_{dos}^{*}$ would be that the calculated m$_{dos}^{*}$ is \textbf{k}-points dependent, whereas the reported experimental results provides the overall value for the materials. However, the values in Table~\ref{table:effmass} are in the range of experimental values. For an excellent {\small{PV}} materials, the effective mass should be low, which corresponds to high mobility of electrons/holes at the CB/VB and resultantly high conductivity. For a material, it is reported that excellent carrier mobility can be obtained by having an effective mass value lower than 0.5m$_{e}$ in at least one direction \cite{ashwin2017tailoring,le2014semiconductors}. In Table~\ref{table:effmass}, at least one high-symmetric point has a m$_{dos}^{*}$ value of less than 0.5m$_{e}$; therefore, both compounds may have great charge mobilities in the vicinities of $\varepsilon_{v}$ and $\varepsilon_{c}$. 
\begin{table}
\caption{Calculated density of states effective mass m$_{dos}^{*}$ \big(units of free electron mass (m$_{e}$)\big) at high-symmetric points for the VB and the CB extrema of Mg$_{2}$Si and Ca$_{2}$Si.}
\resizebox{0.47\textwidth}{!}{
\begin{tabular}{@{\extracolsep{\fill}} c c c c c c c c c c}
 \hline\hline
&& \multicolumn{5}{c}{Values of m$_{dos}^{*}$ for Mg$_{2}$Si/Ca$_{2}$Si compound}  && \\
\cline{3-9} \\ 
High-&& \multicolumn{2}{c}{PBE functional} &  & \multicolumn{2}{c}{mBJ potential} &  & \\
\cline{2-4} \cline{6-9}
symmetric points  &&VB &CB  &&VB &CB && \\
\hline    
W &      &  0.28/0.70  & 0.79/0.60     & & 0.30/1.19       & 0.77/0.57 &&  \\
L &      &  0.51/0.64  & 0.22/0.65     & & 0.56/0.91       & 0.24/0.66 &&  \\
$\Gamma$& & 0.14/0.50  & 0.13/0.20     & & 0.17/0.80       & 0.16/0.64 &&  \\
X &   & $-$0.47/1.17   & 0.27/0.30     & & $-$0.50/1.25    & 0.29/0.41 &&  \\
K &   & $-$0.64/8.80   & 0.39/0.45     & & $-$0.69/$-$6.54 & 0.41/0.51 &&  \\
\\
\hline \hline 
\end{tabular}}
\label{table:effmass}
\end{table}

\subsection{\label{sec:level2}Optical properties}

The operation and performance of solar devices are greatly affected by the optical properties of the underlying compound. Basically, these properties are obtained from the response function calculations in terms of frequency and wave vector dependent complex dielectric function $\epsilon(\omega)$. This has two contributions, which are due to inter-band and intra-band transitions. The contribution by the intra-band transitions is only significant for metals. The inter-band transitions are further classified as direct and indirect transitions. Here, the indirect inter-band transitions, which involve phonon scattering and are expected to contribute not much to $\epsilon(\omega)$ \cite{smith1971photoelectron}, are ignored. In this paper, the imaginary part of the dielectric function $\epsilon_{2}(\omega)$ is calculated at various levels of approximation. First, in IPA by the sum over independent transitions between occupied ($\nu$) and unoccupied ($c$) eigenstates of a one-particle Hamiltonian one defines
\begin{eqnarray} 
    \begin{gathered}
        \epsilon_{2}^{IPA}(\omega) = \frac{Ve^{2}}{2\pi\hbar\omega^{2}m^{2}} \sum_{\nu c} \int \mathrm{d^{3}}\textbf{k}\Big|\big<\nu\textbf{k}\mid\textbf{p}\mid c\textbf{k}\big>\Big|^{2} 
         \\
        \times f(\nu\textbf{k})\big[1-f(c\textbf{k})\big]\delta(\varepsilon_{c\textbf{k}}-\varepsilon_{\nu\textbf{k}}-\hbar\omega)
    \end{gathered}
  \label{eq:ipa}
\end{eqnarray}
where $\hbar\omega$ is the incident photon energy, $V$ is the unit cell volume, \textbf{p} is the momentum operator, and $f(\nu\textbf{k})$ indicates the Fermi-distribution function. The $\delta$ function implies that the absorption occurs at energy eigenvalue differences ($\varepsilon_{c\textbf{k}}$ - $\varepsilon_{\nu\textbf{k}}$). The estimation of matrix elements of momentum operator is performed separately over the muffin-tin and the interstitial regions. A complete description of the evaluation of these matrix elements can be found in the paper of \textit{Draxl et al.} \cite{ambrosch2006linear}. 

\begin{figure*}
\centering
\includegraphics[width=14.0cm, height=7.7cm]{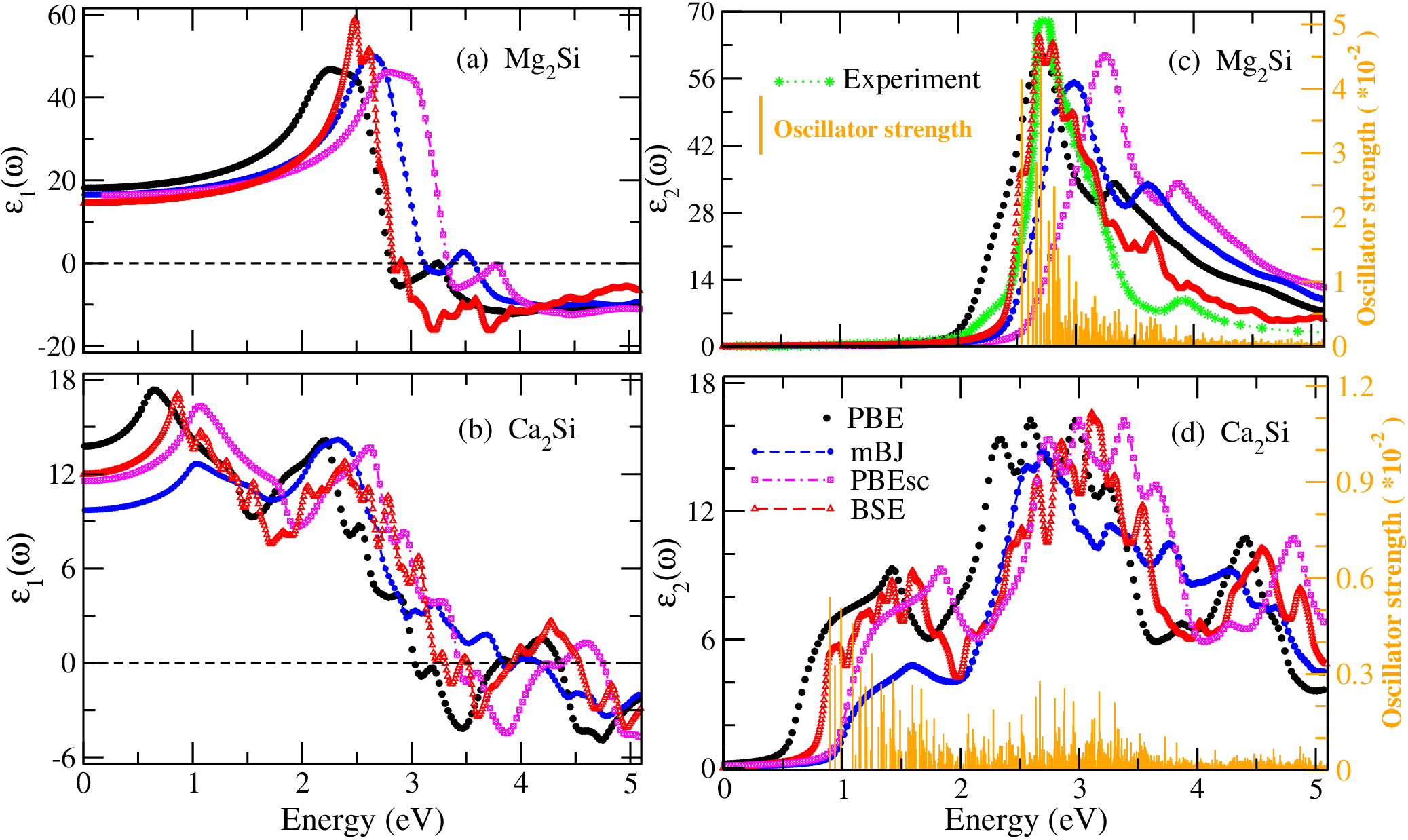} 
\caption{Calculated real $\epsilon_{1}(\omega)$ and imaginary $\epsilon_{2}(\omega)$ parts of the dielectric function from IPA \big(PBE, PBE+scissor (PBEsc), and mBJ\big) and Bethe-Salpeter equation (BSE). In sub-figure \hyperref[fig:epsilon]{(c)}, the experimental spectrum \cite{scouler1969optical} for Mg$_{2}$Si is shown for comparison. The oscillator strength of individual BSE excitations in \hyperref[fig:epsilon]{(c)} \& \hyperref[fig:epsilon]{(d)} is indicated by the vertical orange line.}
\centering
\label{fig:epsilon} 
\end{figure*}

On the other hand, this method does not consider interacting $e$-$h$ pairs and local field effects, which modify the overall shape of the optical dielectric function in a quite significant way. These changes can be made within the BSE approach on the top of DFT band structure from the MBPT \cite{vorwerk2019bethe,onida2002electronic}. In practice, the BSE is mapped onto an eigenvalue problem for an effective two QP excitonic Hamiltonian $H^{BSE}$ \cite{rohlfing2000electron,sagmeister2009time};
\begin{eqnarray}
\sum_{\nu^{\prime}c^{\prime}\textbf{k}^{\prime}} H^{BSE}_{\nu c\textbf{k}, \nu^{\prime}c^{\prime}\textbf{k}^{\prime}} A^{\lambda}_{\nu^{\prime}c^{\prime}\textbf{k}^{\prime}} = E^{\lambda} A^{\lambda}_{\nu c\textbf{k}}                                        
\end{eqnarray}
where $\nu c\textbf{k}$ is an elementary excitation taking place from $\nu$ to $c$ state at a particular \textbf{k}-point in the IBZ. The eigenvalues $E^{\lambda}$ correspond to the energy of coupled $e$-$h$ excitation, whereas the eigenvectors $A^{\lambda}_{\nu c\textbf{k}}$ is the excitation state in terms of coupling between $\lambda^{th}$ $e$-$h$ pair generated by $\nu \textbf{k}\rightarrow c\textbf{k}$ transitions. In two-particle interaction, the $H^{BSE}$ is basically sum of the two terms as follows: 
\begin{eqnarray} \label{eq:}
H^{BSE} = \big( \varepsilon_{c\textbf{k}}-\varepsilon_{\nu\textbf{k}} +\Delta \big) \delta_{cc^{\prime}}\delta_{\nu\nu^{\prime}}\delta_{\textbf{k}\textbf{k}^{\prime}} + \varXi_{\nu c\textbf{k}, \nu^{\prime}c^{\prime}\textbf{k}^{\prime}}      
\end{eqnarray} 
The first term is the diagonal term that describes the transitions within the IPA with QP corrections. In BSE calculations, it is common to adopt the scissor operator to simulate the self-energy (or QP) corrections. Thus, to correct the band gap profile, we simply apply a rigid shift on the CB, known as scissor correction ($\Delta$), which adjusts the optical gap of DFT calculation to mBJ. The second term is known as the effective $e$-$h$ interaction kernel, and written for spin-singlet excitons (when spin is not explicitly treated) \cite{vorwerk2019bethe} as: $\varXi_{\nu c\textbf{k}, \nu^{\prime}c^{\prime}\textbf{k}^{\prime}}$ = 2$\varXi_{\nu c\textbf{k}, \nu^{\prime}c^{\prime}\textbf{k}^{\prime}}^{exc}$ + $\varXi_{\nu c\textbf{k}, \nu^{\prime}c^{\prime}\textbf{k}^{\prime}}^{dir}$. The first one is the short range $exchange$ or unscreened Coulomb repulsion of vanishing wavevector ($\textbf{q}\to 0$) mediated by the bare Coulomb potential. The second one is direct statically $screened$ Coulomb attraction of a negatively charged electron in CB and a positively charged hole in VB \cite{vorwerk2019bethe} and responsible for the creation of the excitons. Within the TDA, the coupling between the resonant and anti-resonant components of $H^{BSE}$ has been excluded \cite{vorwerk2019bethe}. Finally in optical limit ($\textbf{q}\to 0$), the imaginary part of macroscopic dielectric function can be written as;
\begin{eqnarray} \label{eq:exc}
    \begin{gathered}
        \epsilon_{2}^{BSE}(\omega) = \lim_{\textbf{q}\to 0}\frac{8\pi^{2}}{\Omega q^{2}}\sum_{\lambda}\Bigg|\sum_{\nu c\textbf{k}} A^{\lambda}_{\nu c\textbf{k}}\big<\nu\textbf{k}\mid     e^{-\iota\textbf{q}.\textbf{r}}\mid c\textbf{k}\big>\Bigg|^{2} 
         \\
        \times \delta(E^{\lambda}-\omega)
    \end{gathered}
\end{eqnarray}
where $\Omega$ is the crystal volume. The square modulus of coupling of different $e$-$h$ pair configurations $|\nu c\big>$ in excitation state $\lambda$ is known to be the oscillator strength in the optical spectrum. Equation~\hyperref[eq:exc]{\eqref{eq:exc}} should be compared with Eq.~\hyperref[eq:ipa]{\eqref{eq:ipa}}; the differences between them are attributed to change in excitation energies (from ($\varepsilon_{c}$ - $\varepsilon_{\nu}$) to $E^{\lambda}$), as well as, importantly, the coefficients $A^{\lambda}$, which enable the interpretation of spectra in terms of the mixing of formerly independent transitions.

The real part of dielectric function $\epsilon_{1}(\omega)$ can be obtained from the Kramers-Kronig ($KK$) relation \cite{ambrosch2006linear,solet2022first} as:
\begin{eqnarray} \label{eq:eps1}
\epsilon_{1}(\omega) = 1+ \frac{2}{\pi} \int_{0}^{\infty} \frac{\epsilon_{2}(\omega^{\prime})\omega^{\prime}d\omega^{\prime}}{\omega^{\prime 2}-\omega^{2}}                                       
\end{eqnarray}
For a good accuracy of $\epsilon_{1}(\omega)$ calculation, one needs to take good representation of $\epsilon_{2}(\omega)$ up to high energies. In IPA calculation, $\epsilon_{2}(\omega)$ is calculated up to 100 eV and also this value used as a truncation energy in Eq.~\hyperref[eq:eps1]{\eqref{eq:eps1}}. Finally, the other optical parameters such as refractive index, absorption coefficient, and reflectivity can be easily obtained \cite{ambrosch2006linear,solet2022first}.

It is important to note that the thermoelectric transport properties can be explained in better way by considering the band-gap correction on the different {\small{XC}} functionals  \cite{sk2018exploring,shastri2020first,shastri2020thermoelectric}. Both thermoelectric and PV transport properties are highly dependent upon electronic band structures. Thus, a reliable result of PV properties can be obtained with the scissor correction value on PBE functional. This value for Mg$_{2}$Si and Ca$_{2}$Si is used to be 0.55 eV and 0.4 eV, respectively, which is the difference between the optical band gap of mBJ and PBE calculations.

Therefore, in this section, we discuss the optical spectra of both silicides using different methods within IPA (PBE, PBE with scissor (PBEsc), and mBJ). The dielectric function, in Fig.~\ref{fig:epsilon}, estimated within the IPA is compared to the ones considering many-body excitonic effects by solving the BSE. In Fig.~\hyperref[fig:epsilon]{\ref{fig:epsilon}(a)}, the static electronic dielectric function $\epsilon_{1}(0)$ (or $\epsilon_{\infty}$) from PBE is obtained to be $\sim$18 for Mg$_{2}$Si, which overestimates the experimental value of $\sim$13 \cite{scouler1969optical}. Additionally, the $\epsilon_{1}(0)$ from PBEsc ($\sim$15.6) and mBJ ($\sim$16.5) fails to address this overestimating tendency. However, the BSE method estimates this value of $\sim$14.4, which is more consistent with the experimental value. This macroscopic electronic $\epsilon_{\infty}$ can be influenced by considering the approximation beyond TDA \cite{bechstedt2016many}. Further, the highest peak of $\epsilon_{1}(\omega)$ is observed to be in the visible energy region of $\sim$2.5 eV. In Fig.~\hyperref[fig:epsilon]{\ref{fig:epsilon}(b)}, the obtained $\epsilon_{1}(0)$ values from PBE, PBEsc, and mBJ are $\sim$13.8, 11.5, and 9.7, respectively, for Ca$_{2}$Si. Including the $e$-$h$ interaction in the PBE reduces the $\epsilon_{1}(0)$ to $\sim$12. In the ultraviolet (UV) energy region, the majority of the $\epsilon_{1}(\omega)$ is negative, which means both materials show metallic character in this energy spectrum \cite{ashcroft1976}.

\begin{figure*}
\centering
\includegraphics[width=14.1cm, height=7.3cm]{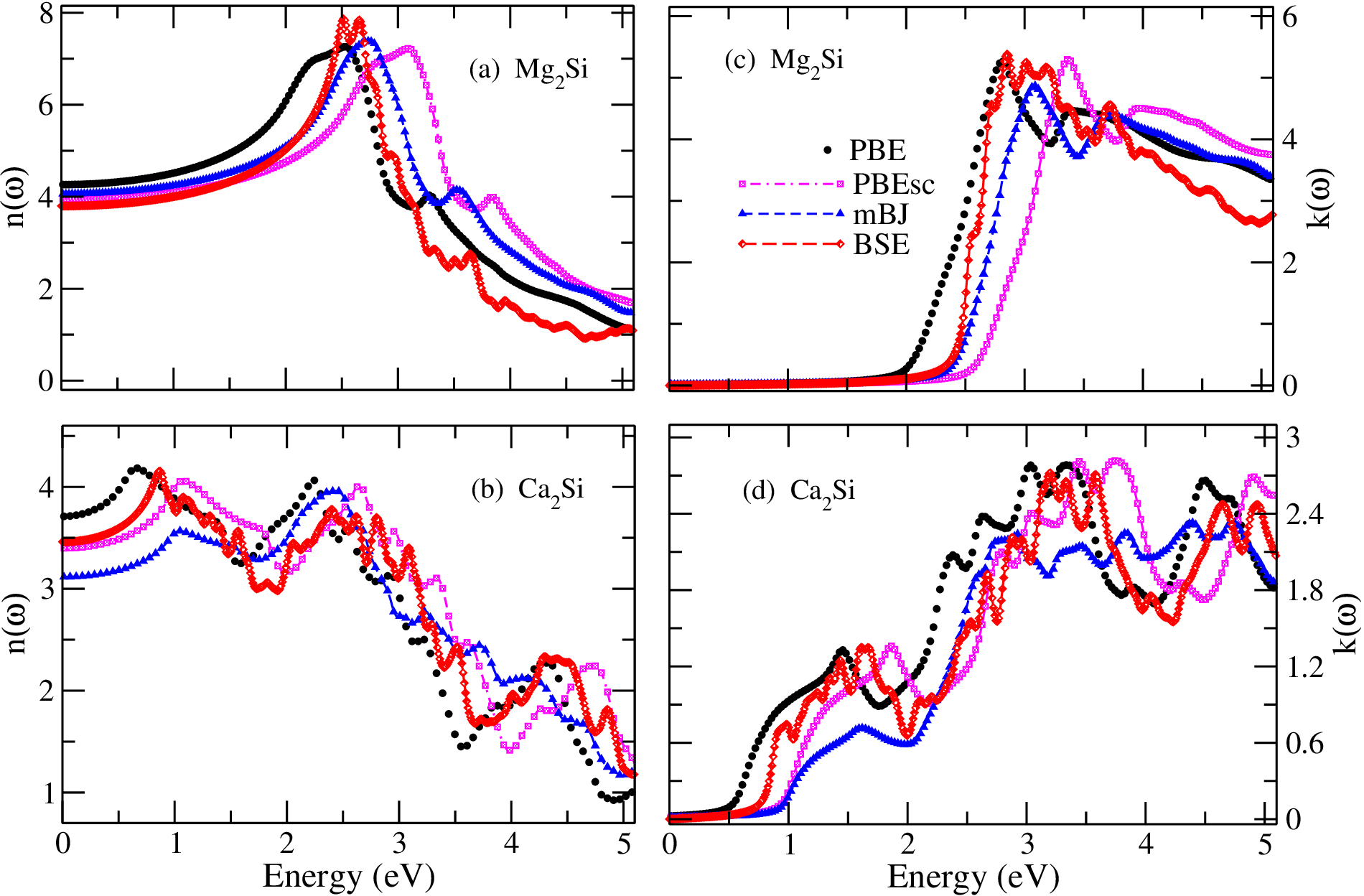} 
\caption{The real n$(\omega)$ and imaginary k$(\omega)$ parts of refractive index comparing IPA \big(PBE, PBE+scissor (PBEsc), mBJ\big) with Bethe-Salpeter equation (BSE) for Mg$_{2}$Si and Ca$_{2}$Si compounds.}
\centering
\label{fig:refra} 
\end{figure*}

In Fig.~\hyperref[fig:epsilon]{\ref{fig:epsilon}(c)}, the calculated $\epsilon_{2}(\omega)$ spectra together with the available experimental data by Scouler \cite{scouler1969optical} from Reflectance measurement on Mg$_{2}$Si are presented. The tail that appears below the onset of respective absorption in both IPA and BSE is caused by the lifetime broadening effects, which resemble the self-energy contribution. The $\epsilon_{2}(\omega)$ obtained within the IPA with different methods reproduces mainly two bright peaks. The maximum of the first main peak appears at $\sim$2.7, 3.3, and 3 eV from the PBE, PBEsc, and mBJ, respectively. For PBE, this peak nearly coincides with the first peak in the experimental $\epsilon_{2}(\omega)$ at $\sim$2.73 eV. The spectra obtained from PBEsc show only a rigid shift relative to the PBE spectra, while the spectra within mBJ almost lie between the two. However, none of the three methods accurately reproduce spectra that can be compared with experimental results, as the oscillator strengths are not represented accurately. In Fig.~\hyperref[fig:epsilon]{\ref{fig:epsilon}(d)}, Ca$_{2}$Si exhibits multiple peaks of $\epsilon_{2}(\omega)$ from all methods, revealing several optical transitions for the particular dispersion of its $\varepsilon_{v}$ and $\varepsilon_{c}$. In this figure, the $\epsilon_{2}(\omega)$ is increasing slowly from respective absorption edges to the first peak in all three IPA as well as BSE methods. All methods produce the highest peak in energy range of $\sim$2.6-3.4 eV. One can see that the BSE calculation gives the red-shift on the PBEsc $\epsilon_{2}(\omega)$ spectrum, but the intensities of peaks are almost unchanged. The highest $\epsilon_{2}(\omega)$ within BSE lies in the visible energy region, which is $\sim$16.7 at 2.1 eV. Until now, experimental optical spectra of Ca$_{2}$Si have not been available. On the other hand, in Fig.~\hyperref[fig:epsilon]{\ref{fig:epsilon}(c)}, the $\epsilon_{2}(\omega)$ spectrum calculated within BSE shifts towards lower energies compared to the PBEsc/mBJ spectrum. The shape of the spectrum is now improved and better described compared to the experiment, with a redistribution of transitions relative to the IPA results. However the BSE curve does not match the experimental data below the direct gap energy (2.45 eV) due to the exclusion of indirect transitions in the calculation. It shows fairly good agreement with the experimental counterpart at $\sim$2.73 eV, with the calculated magnitude for BSE being $\sim$65 compared to the experimental value of 68. A large excitonic effect is observed in Mg$_{2}$Si than Ca$_{2}$Si, particularly at the optical gap edge. This may be due to the large optical band gap of Mg$_{2}$Si than Ca$_{2}$Si. Above $\sim$3.3 eV in the UV energy region, there is a notable discrepancy between the BSE and experimental spectra. This may be due to the current calculations not considering the off-diagonal elements in the screening matrix \cite{sponza2013role}, frequency-dependent (dynamical) screening \cite{zhang2023effect}, and electron-phonon coupling or polaronic excitation effects \cite{devreese2010many}. The well amplified energy peak from BSE at 2.54 eV is almost concealed in the Mg$_{2}$Si experimental spectrum. This could be due to two factors: a lower broadening value of 50 meV used in calculation, or the technique used to extract the experimental spectrum. In the latter case, Scouler \cite{scouler1969optical} deduced the $\epsilon_{2}(\omega)$ from $KK$ analysis of the reflectance data, which was measured only up to 11 eV. To perform a much more accurate $KK$ transform, one must use the reflectance spectrum up to high photon energies. To produce a high-energy portion of the reflectance data, Scouler used a parameter-based tail function. However, he suggested that the positions and magnitudes of peaks can influence the choice of tail function \cite{scouler1969optical}, which complicates direct comparisons with theoretical results.

\begin{figure*}
\centering
\includegraphics[width=14.1cm, height=7.2cm]{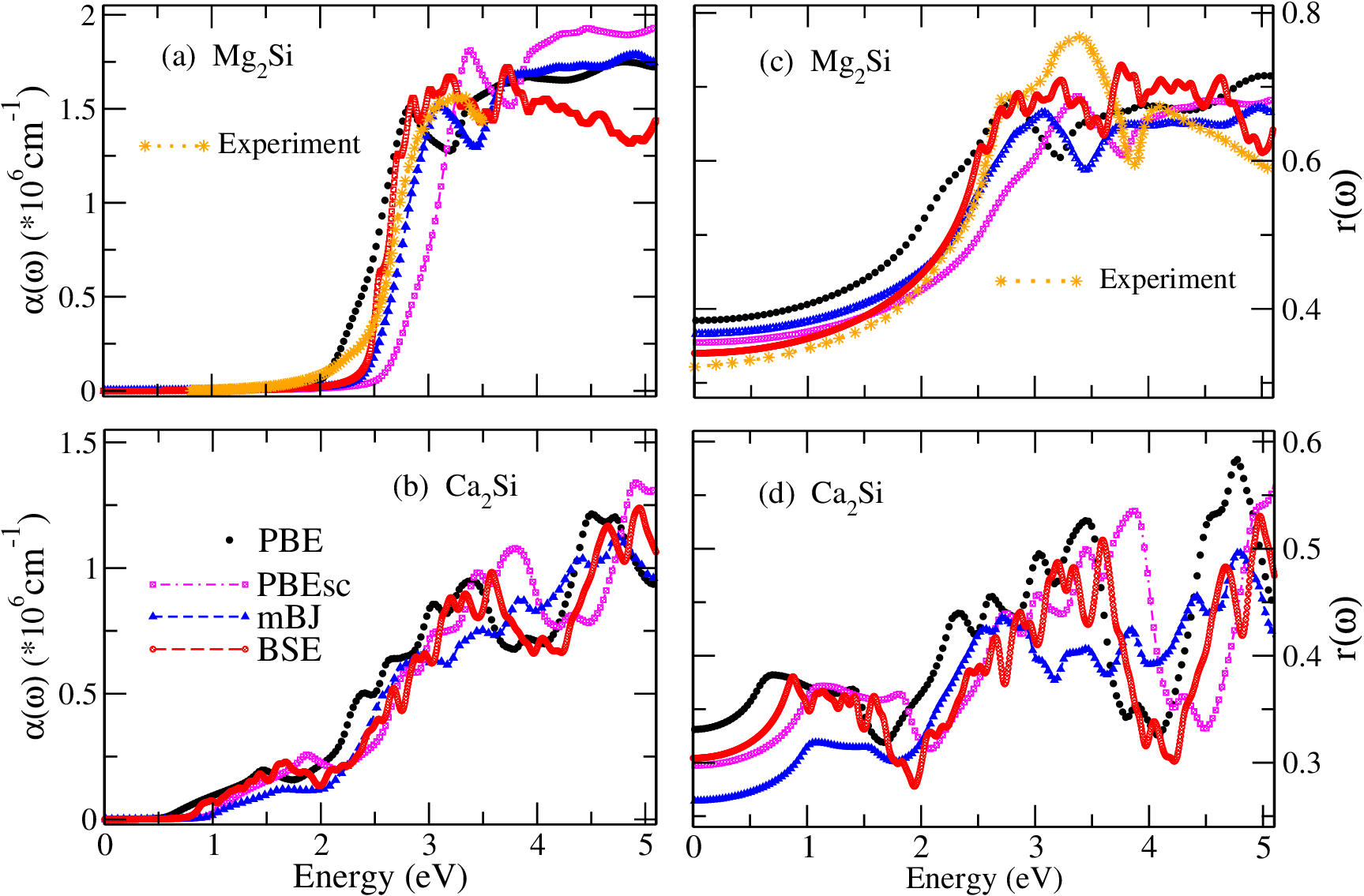} 
\caption{Calculated absorption coefficient $\alpha(\omega)$ and reflectivity r$(\omega)$ for Mg$_{2}$Si and Ca$_{2}$Si: a comparison of IPA methods \big(PBE, PBE+scissor (PBEsc), and mBJ\big) with the Bethe-Salpeter equation (BSE). The experimental data in panels \hyperref[fig:abs_rfl]{(a)} and \hyperref[fig:abs_rfl]{(c)} are from references \cite{kato2011optoelectronic} and \cite{scouler1969optical}, respectively.}
\centering
\label{fig:abs_rfl} 
\end{figure*}

In order to gain a deeper insight into the origin of excitonic peaks that contribute to the $\epsilon_{2}(\omega)$, we have plotted the oscillator strength of excitons for both silicides in Figs.~\hyperref[fig:epsilon]{\ref{fig:epsilon}(c)} and ~\hyperref[fig:epsilon]{\ref{fig:epsilon}(d)}. If it is negligibly small, exciton is known to be dark; otherwise it is considered bright for its contribution to the spectrum. The strength of the excitons in Mg$_{2}$Si is comparably larger than those for Ca$_{2}$Si. The first bright exciton in Fig.~\hyperref[fig:epsilon]{\ref{fig:epsilon}(c)} is estimated at $\sim$2.54 eV with an intensity of $\sim$0.041, making a contribution to the formation of a little hump in the Mg$_{2}$Si $\epsilon_{2}(\omega)$ spectrum. The second significant bright exciton, seen at $\sim$2.7 eV, originates the highest peak comparable to the experiment in the BSE spectrum. For Ca$_{2}$Si, the only bound bright exciton in Fig.~\hyperref[fig:epsilon]{\ref{fig:epsilon}(d)} appears at $\sim$0.89 eV, which is below the direct gap of $\sim$0.96 eV. Its binding energy of only 70 meV indicates that this exciton is loosely bound. This may be attributed to the lower band gap as well as lower m$_{dos}^{*}$ value of electron at $\varepsilon_{c}$ \cite{phillips1966fundamental,ashcroft1976} (see Table~\ref{table:effmass}) in Ca$_{2}$Si. So, this bound exciton can easily dissociate into free carriers with the aid of heat energy or additional photons. This can enhance charge carrier generation in Ca$_{2}$Si, making it more efficient for applications like solar cells. After an optical energy region, both silicides have a large number of dark excitons and may have an insignificant contribution of effective two QP excitons at static screening to the $\epsilon_{2}(\omega)$ in quantitative agreement with experiment.

Fig.~\ref{fig:refra} illustrates the computed real n$(\omega)$ and imaginary k$(\omega)$ parts of the refractive index for both silicides using IPA and BSE. The n$(\omega)$ \big(k$(\omega)$\big) follows the same trend as observed in the $\epsilon_{1}(\omega)$ \big($\epsilon_{2}(\omega)$\big) plot for the respective compounds. The range of n(0) is observed to be $\sim$3.8$-$4.26 for Mg$_{2}$Si and 3.1$-$3.7 for Ca$_{2}$Si. This can be attributed to the inverse relationship between refractive index and band gap, since the band gap of the former is lower than that of the latter (see Table~\ref{table:bandgap}) \cite{tripathy2015refractive}. The n(0) value from BSE of $\sim$3.8 is found to be very close to the experimental n(0) of $\sim$3.6 \cite{scouler1969optical} for Mg$_{2}$Si. Also, the maximum n$(\omega)$ using BSE of $\sim$7.9 at $\sim$2.5 eV is fairly good agreement with the experimental magnitude of $\sim$7.7 at $\sim$2.55 eV. Next, the range of maximum intensity of k$(\omega)$ for Mg$_{2}$Si and Ca$_{2}$Si is found to be $\sim$2.6$-$4.0 eV and $\sim$2.4$-$4.0 eV, respectively. This implies that in this energy range, photons will be absorbed very quickly (i.e., the penetration depth will be the shortest within this range). The highest k$(\omega)$ of $\sim$5.4 at $\sim$2.85 eV photon energy from BSE has the best matching with experimental peak of 5.4 at $\sim$2.95 eV \cite{scouler1969optical}. Overall, both n$(\omega)$ and k$(\omega)$ have larger values for Mg$_{2}$Si than Ca$_{2}$Si in the studied energy range. Due to the large n$(\omega)$ ($\sim>$3) of both silicides, they can be the best materials to design the anti-reflection coating surface on solar devices \cite{trupke2002improving}. 

We compare in Fig.~\ref{fig:abs_rfl} the absorption coefficient $\alpha(\omega)$ and reflectivity r$(\omega)$ from IPA with BSE calculations. The $\alpha(\omega)$ from spectroscopic ellipsometry measurement by \textit{Kato et al.} \cite{kato2011optoelectronic} on poly-crystalline Mg$_{2}$Si thin-film is also plotted for comparison in Fig.~\hyperref[fig:abs_rfl]{\ref{fig:abs_rfl}(a)}. In this figure, the red-shift of the PBE $\alpha(\omega)$ has become the blue-shift after applying the scissor correction on PBE (PBEsc) with respect to the experiment. Additionally, the presence of $e$-$h$ interaction on PBEsc significantly improves the $\alpha(\omega)$ and aligns well with the experiment. Beyond the optical gap edge ($\sim$2.45 eV), the agreement of $\alpha(\omega)$ (from BSE) with the experiment is satisfactory, and below the gap it is not matching due to consideration of direct optical transitions only. One can also notice that the BSE $\alpha(\omega)$ increases rapidly from an optical gap onset to an edge of the visible energy region of $\sim$3.2 eV and then decreases as oscillatory type in the UV energy region. In Fig.~\hyperref[fig:abs_rfl]{\ref{fig:abs_rfl}(b)}, the $\alpha(\omega)$ from all IPA and BSE methods has been increased monotonically with photon energy. Also, we do not observe a significant change, especially up to visible light, in the BSE spectrum compared to the QP-corrected PBEsc spectrum, which means two interacting QP picture may be less effective in Ca$_{2}$Si. The highest intense peak in the visible energy regime (at $\sim$3.2 eV) is found to be $\sim$1.7 and $\sim$0.9 $\times$$10^{6}$ cm$^{-1}$ for Mg$_{2}$Si and Ca$_{2}$Si, respectively. Further, for Mg$_{2}$Si, the BSE calculation has provided a comparable r$(\omega)$ up to visible energy region to the Scouler reflectance measurement \cite{scouler1969optical} in Fig.~\hyperref[fig:abs_rfl]{\ref{fig:abs_rfl}(c)}. In the considered UV light region, the r$(\omega)$ underestimates (overestimates) experimental data for photon energies below (above) $\sim$3.7 eV. The minimum r$(\omega)$ for visible light from BSE is estimated to be $\sim$41\%. The spectrum from all methods initially shows a linear increase in behaviour (same as experiment) up to their highest value in the visible energy region, and after that, the spectrum does not show too many changes with energy. In Fig.~\hyperref[fig:abs_rfl]{\ref{fig:abs_rfl}(d)}, the r$(\omega)$ in Ca$_{2}$Si exhibits a more of an oscillatory nature within the studied energy window across all methods. In the infrared (IR) region, the mBJ method provides the lowest r$(\omega)$ among all the methods. Beyond this region, the spectrum with excitonic contributions in PBEsc (BSE) yield the lowest values, particularly in the visible light, reaching a minimum of $\sim$27\%. This low reflectivity suggests that Ca$_{2}$Si could be used to make an anti-reflective coating surface.

The optical properties obtained for Mg$_{2}$Si are much more comparable to those of silicon, while the properties for Ca$_{2}$Si show a better match with various solar compounds such as InAs, GaAs, and GaP \cite{philipp1963optical}. However, possessing excellent optical properties is not a sufficient criterion for using materials in PV technology. Therefore, it is essential to examine the material’s power conversion efficiency to determine its suitability for solar cells.

\subsection{\label{sec:level2}{\small{SLME}} and defect-aided {\small{SRH}} mechanisms}
The light to electricity conversion efficiency from solar cells depends on the extraction of $e$-$h$ pairs by materials and is limited by the various $e$-$h$ recombination processes. In this paper, {\small{SLME}} of single-junctional {\small{PV}} cell is calculated from current-voltage analysis, explicitly focusing on radiative recombination \cite{yu2012identification}. While, the trap-limited conversion ({\small{TLC}}) efficiency $\eta_{{\small{TLC}}}$ is estimated by considering both radiative and non-radiative recombination mechanisms \cite{nelson2003the, kim2020upper, kirchartz2018makes}. We mainly consider the optoelectronic and point-defect properties of materials. Let us first discuss the radiative limit. Following the Shockley diode equation, the radiative recombination current density $J_{rad}$ for net voltage across the cell ($V$) can be estimated by \cite{kirchartz2018makes},  
\begin{eqnarray} \label{eq:jr}
J_{rad} = J_{rad}^{0}\Bigl(e^{(eV/ k_{B}T)} - 1\Bigl),
\end{eqnarray}
and $J_{rad}^{0}$ is due to black-body photons at temperature $T$ absorbed by the front cell surface from the surrounding thermal bath
\begin{eqnarray} \label{eq:j0r}
J^{0}_{rad} = e\pi\int^{\infty}_{0}\Bigl(1 - e^{-2\alpha(\varepsilon)L}\Bigl)I_{bb}(\varepsilon, T)d\varepsilon,
\end{eqnarray}
with $I_{bb}(\varepsilon, T)$ is the black-body radiance related to the radiative process, $e$ is the elementary charge and $k_{B}$ is the Boltzmann constant. $\alpha(\varepsilon)$ is material absorption coefficient and $L$ is the thickness of an absorber layer with zero and unity-reflectivity from the front and back surface (hence the presence of factor 2), respectively.

In the case of non-radiative recombination, $e$-$h$ pairs recombine through trap states within the band gap, as described by the {\small{SRH}} model \cite{shockley1952statistics}. In the high-injection limit where both charge carriers are present in almost similar concentration, the non-radiative current density is \cite{kirchartz2018makes};
\begin{eqnarray} \label{eq:jnr}
    \begin{gathered}
        J_{nr} = J_{nr}^{0}\Bigl(e^{(eV/ 2k_{B}T)} - 1\Bigl), 
         \\
         J^{0}_{nr} = \frac{eL\sqrt{N_{c}N_{v}}}{2\uptau} e^{-\varepsilon_{g}/2k_{B}T}, 
         \\
         N_{c} = \int^{\infty}_{\varepsilon_{c}} g_{c}(\varepsilon) e^{\frac{-(\varepsilon-\varepsilon_{c})}{k_{B}T}} d\varepsilon, \quad N_{v} = \int^{\varepsilon_{v}}_{-\infty}  g_{v}(\varepsilon) e^{\frac{-(\varepsilon_{v}-\varepsilon)}{k_{B}T}} d\varepsilon . 
    \end{gathered}
\end{eqnarray}
where $N_{c}$ ($N_{v}$) and $g_{c}(\varepsilon)$ ($g_{v}(\varepsilon)$) are an effective and total density of states in the CB (VB), respectively \cite{ashcroft1976}. $\varepsilon_{g}$ is the band gap of an absorber and $\uptau$ is the nonradiative recombination lifetime (or {\small{SRH}} lifetime).

Finally the trap-limited current density $J$ of an illuminating solar cell under photon flux at bias voltage $V$ is given by,
\begin{eqnarray} \label{eq:j}
J(V; L) = J_{sc}(L) - J_{rad}(V; L) - J_{nr}(V; L).
\end{eqnarray}

The first term is the short-circuit current density and defined as,
\begin{eqnarray} \label{eq:jsc}
J_{sc} = e\int^{\infty}_{0} \Bigl(1 - e^{-2\alpha(\varepsilon)L}\Bigl) I_{sun}(\varepsilon) d\varepsilon,
\end{eqnarray}
where $I_{sun}(\varepsilon)$ is the incident total number of photons of energy $\varepsilon$ per unit area, energy and time, and standard AM1.5G flat-plate solar spectrum at 25 \textcelsius \cite{AM1.5} is used for this. Finally, the $\eta_{{\small{TLC}}}$ from maximum output power is defined as,
\begin{eqnarray} \label{eq:tce}
\eta_{{\small{TLC}}} = \frac{max(J\times V)}{\int^{\infty}_{0}\varepsilon I_{sun}(\varepsilon)d\varepsilon} .
\end{eqnarray}

In {\small{SLME}} formalism \cite{yu2012identification,solet2022first}, for calculating $J$, only first two terms in right side of Eq.~\hyperref[eq:j]{\eqref{eq:j}} are considered. The extra factor $e^{\Delta/k_{B}T}$, which is the fraction of the radiative recombination current, is also multiplied in Eq.~\hyperref[eq:jr]{\eqref{eq:jr}}. The $\Delta$ = $\varepsilon_{g}^{da} - \varepsilon_{g}$, where $\varepsilon_{g}^{da}$ is direct-allowed band gap of material. Then the estimated $J$ is used in Eq.~\hyperref[eq:tce]{\eqref{eq:tce}} to calculate {\small{SLME}}.

Next, the {\small{TLC}} efficiency of PV solar cell can be calculated by solving Eq.~\hyperref[eq:tce]{\eqref{eq:tce}} after information on electronic dispersion, carrier recombination lifetime $\uptau$, absorption coefficient, and the thickness $L$ of an absorbing layer. Now let us develop an expression of $\uptau$. As we have already discussed in the introduction that, the {\small{SRH}} model defines the recombination rate of $e$-$h$ pairs via a trapping state inside the band gap \cite{shockley1952statistics}. Based on the principle of detailed balance, the {\small{SRH}} recombination rate ($R_{{\small{SRH}}}$) of electrons in $\varepsilon_{c}$ and holes in $\varepsilon_{v}$ associated to a defect density $N_{d}$ of a single trap state $t$ at an energy $\varepsilon_{t}$ is given by \cite{shockley1952statistics,hall1952electron},
\begin{eqnarray} \label{eq:rt}
R_{{\small{SRH}}} = \frac{np - n_{0}p_{0}}{\uptau_{n0} (n + n_{t}) + \uptau_{p0} (p + p_{t})},
\end{eqnarray} 
where,
\begin{eqnarray} \label{eq:nt}
    \begin{gathered}
        \uptau_{n0} = \frac{1}{N_{d}C_{n}}, \quad \uptau_{p0} = \frac{1}{N_{d}C_{p}},
         \\
        n_{t} = N_{c} e^{-(\varepsilon_{c} - \varepsilon_{t})/k_{B}T}, \quad p_{t} = N_{v} e^{-(\varepsilon_{t} - \varepsilon_{v} )/k_{B}T}. 
    \end{gathered}
\end{eqnarray}
Here $\uptau_{n0}$, $C_{n}$ and $\uptau_{p0}$, $C_{p}$ are the capture time constants, capture coefficients for an electron in CB and a hole in VB at trap state $t$, respectively. After that, $n_{t}$ and $p_{t}$ are the number of electrons and holes for the case in which $\varepsilon_{F}$ falls at $\varepsilon_{t}$.

Accordingly, the steady-state density of electrons $n$ and holes $p$ under applied voltage $V$ (or under illumination) for Eq.~\hyperref[eq:rt]{\eqref{eq:rt}} can be estimated by:
\begin{eqnarray} \label{eq:np}
n = n_{0} + \Delta n, \quad and \quad p = p_{0} + \Delta n.
\end{eqnarray}

Finally, the carrier lifetime associated to the recombination rate $R_{{\small{SRH}}}$ with excess carrier densities $\Delta n$ generated from illumination is given by,
\begin{eqnarray} \label{eq:tau}
\uptau = \frac{\Delta n}{R_{{\small{SRH}}}}.
\end{eqnarray}

Now, the unknown quantities to calculate $\uptau$ in Eq.~\hyperref[eq:tau]{\eqref{eq:tau}} are basically concentrations ($n_{0}$, $p_{0}$, $\Delta n$, and $N_{d}$), $\varepsilon_{t}$ and $C_{n/p}$, which can be estimated from the point defect calculations. The main quantity, which is needed to define the point defect in a system is its formation energy. From the supercell approach, the formation energy $\Delta E_{f}(d^{q})$ of any point defect $d$ in charge state $q$ can be written as follows \cite{alkauskas2014first};
\begin{eqnarray} \label{eq:fe}
    \begin{gathered}
     \Delta E_{f}(d^{q}) = G(d)-G(bulk)-\sum_{i}n_{i}\mu_{i} 
     \\
      + q(\varepsilon_{v} + \varepsilon_{F}). 
    \end{gathered}
\end{eqnarray} 
where $G(d)$ and $G(bulk)$ are the free energies of defect containing and pristine supercells, respectively. The integer $n_{i}$ defines the number of $i$ type atoms which have been added to ($n_{i} > 0$) or removed from ($n_{i} < 0$) the pristine supercell to make a defective system, and $\mu_{i}$ are the respective chemical potentials of the defect species.

We shall now calculate the equilibrium concentrations of a defect and carriers. Typically, we need to know the $\varepsilon_{F}$ to obtain these concentrations. For this purpose, we have used a Fortran code named \texttt{SC-FERMI} \cite{buckeridge2019equilibrium}, which employs an iterative search algorithm to calculate the $\varepsilon_{F}$ that satisfies the charge neutrality constraint for the overall system. This value of $\varepsilon_{F}$ is termed the \textquotedblleft self-consistent Fermi energy\textquotedblright, and used to calculate $\Delta E_{f}(d^{q})$ in Eq.~\hyperref[eq:fe]{\eqref{eq:fe}}. Then the intrinsic electron, hole ($n_{0}$, $p_{0}$) concentrations without illuminating light, as well as the equilibrium concentration of a defect $N_{d}(q)$ for $q$, are calculated as follows \cite{ashcroft1976,buckeridge2019equilibrium}: 
\begin{eqnarray} \label{eq:concent}
    \begin{gathered}
        n_{0} = N_{c} e^{-(\varepsilon_{c} - \varepsilon_{F} )/k_{B}T}, \quad p_{0} = N_{v} e^{-(\varepsilon_{F} - \varepsilon_{v} )/k_{B}T},
         \\
        N_{d}(q) = N_{site}g e^{-\Delta E_{f}(d^{q}) /k_{B}T}. 
    \end{gathered}
\end{eqnarray}
where, $N_{site}$ and $g$ represent the density of available crystal sites and the degeneracy for defect $d$, respectively. For simplicity, we set $g$ to 1.

Based on charge neutrality condition for defects and carriers, one can write \cite{buckeridge2019equilibrium,ashcroft1976},
\begin{eqnarray} \label{eq:fermi}
\sum_{q, d}q N_{d}(q) = p_{0} - n_{0}.
\end{eqnarray}
The dependence of each term in Eq.~\hyperref[eq:fermi]{\eqref{eq:fermi}} on $\varepsilon_{F}$ allows for the use of a search algorithm to iteratively update $\varepsilon_{F}$ until it satisfies the requirements of Eq.~\hyperref[eq:fermi]{\eqref{eq:fermi}} at a given temperature.

The defect formation energies (DFE) of both compounds for two point vacancy defects ($V_{\small{Mg/Ca}}$ and $V_{Si}$), two interstitial defects ($I_{Mg/Ca}$ and $I_{Si}$), and anti-site defects ($Mg/Ca_{Si}$ and $Si_{Mg/Ca}$) have been calculated at $q = 0$ and tabulated in Table~\ref{table:fe}. The DFE of these native defects are obtained in growth conditions rich in $Si$, where $\mu_{Si}$=$E(Si)$, and $\mu_{Mg}$=(1/2)$(\small{E(\small{Mg_{2}Si})-E(Si)})$. The $E(Mg_{2}Si)$ and $E(Si)$ are energies of a bulk stable Mg$_{2}$Si and $Si$ in FCC structures, respectively. The order of calculated DFE from Table~\ref{table:fe} is: $I_{Mg } < Mg_{Si} < V_{Mg} < V_{Si} < I_{Si} < V_{MgSi} < Si_{Mg}$ for Mg$_{2}$Si and $I_{Si} < Si_{Ca} < I_{Ca} <  Ca_{Si} < V_{Si} < V_{Ca} < V_{CaSi}$ for Ca$_{2}$Si. Due to low DFE, the interstitial Mg ($I_{Mg}$) and $Si$ ($I_{Si}$) atoms on 4$b$ sites are the most favorable/stable defects compared to other defects in Mg$_{2}$Si and Ca$_{2}$Si, respectively. The next most stable defects is due to anti-site with $Mg_{Si}$ and $Si_{Ca}$ in the both respective silicides. Thus, we will further calculate $R_{{\small{SRH}}}$ and $\uptau$ by considering only a single local single-particle defect state, which is due to $I_{Mg}$ ($I_{Si}$) interstitial defect in the pristine structure of Mg$_{2}$Si (Ca$_{2}$Si).

\begin{table}[h]
\caption{The obtained defect formation energies (DFE) from PBE for different types of point defects on Mg$_{2}$Si and Ca$_{2}$Si in Si-rich condition.}
\resizebox{0.5\textwidth}{!}{
\begin{tabular}{@{\extracolsep{\fill}} c c c | c c c c c c c c c c c c}
\hline\hline 
Defects in&& \multicolumn{1}{c |}{DFE} &\multicolumn{6}{c}{Defects in}  & \multicolumn{2}{c}{DFE} &  & \\
Mg$_{2}$Si  &&(eV/atom) & &&&Ca$_{2}$Si &&&(eV/atom) && \\
\hline   
$V_{Mg}$     & &  1.50    &&& &$V_{Ca}$    & &&  2.37    &&  \\
$V_{Si}$     & &  2.23    &&& &$V_{Si}$    & &&  2.00    &&  \\
$Mg_{Si}$    & &  1.47    &&& &$Ca_{Si}$   & &&  1.91    &&  \\
$Si_{Mg}$    & &  2.90    &&& &$Si_{Ca}$   & &&  1.53    &&  \\
$I_{Si}$     & &  2.44    &&& &$I_{Si}$    & &&  1.24    &&  \\
$I_{Mg}$     & &  1.35    &&& &$I_{Ca}$    & &&  1.65    &&  \\
$V_{MgSi}$   & &  2.66    &&& &$V_{CaSi}$  & &&  3.12    &&  \\

\hline \hline 
\end{tabular}}
\label{table:fe}
\end{table}

\begin{figure}
\centering
\includegraphics[width=8.3cm, height=8.8cm]{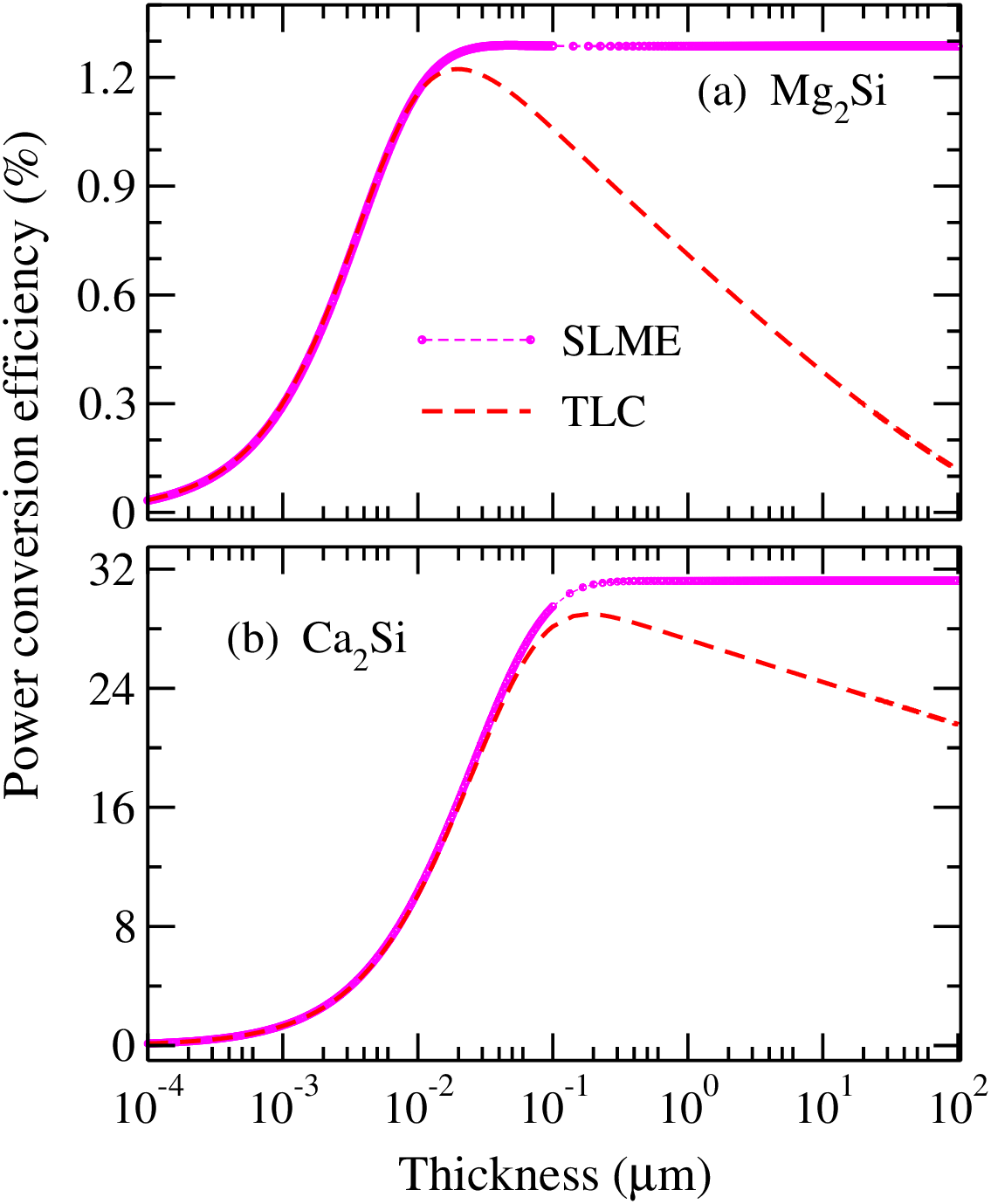} 
\caption{Spectroscopic limited maximum efficiency ({\small{SLME}}) and trap-limited conversion ({\small{TLC}}) efficiency using absorption coefficient from Bethe-Salpeter equation (BSE) method for (a) Mg$_{2}$Si (b) Ca$_{2}$Si as a function of thickness.}
\centering
\label{fig:tlc} 
\end{figure}

\begin{table*}
\caption{Photovoltaic performance parameters of Mg$_{2}$Si and Ca$_{2}$Si using the Bethe-Salpeter equation (BSE) based absorption coefficient at 300 K, predicted by spectroscopic limited maximum efficiency ({\small{SLME}}), Shockley–Queisser (SQ) limit, and trap-limited conversion efficiency ($\eta_{{\small{TLC}}}$). $J_{sc}$ is the short-circuit current density and $V_{oc}$ is the open-circuit voltage.}
\resizebox{0.79\textwidth}{!}{
\begin{tabular}{@{\extracolsep{\fill}} c c c c c c c c c c c c c c c c c c c c c c c c c c c c c c c c c}
\hline\hline 
Solar materials &&& \multicolumn{2}{c}{SRH lifetime} &&\multicolumn{6}{c}{$J_{sc}(mAcm^{-2})$} &&\multicolumn{3}{c}{$V_{oc}(V)$}  && & \multicolumn{3}{c}{$\eta_{{\small{TLC}}}$} &&&\multicolumn{3}{c}{{\small{SLME}}} &&& \multicolumn{3}{c}{SQ-limit}  & \\
\hline \\  
Mg$_{2}$Si    &&& 2   $\mu s$    &&&&&&& 6.8    &&&& 0.55     &&&&&& 1.2\%     &&&& 1.3\%     &&&&& 19.37\% \\
\\Ca$_{2}$Si  &&& 11.3  $ms$    &&&&&&& 50.5   &&&& 0.7      &&&&&& 28.5\%    &&&& 31.2\%    &&&&& 31.25\% \\
\\
\hline \hline 
\end{tabular}}
\label{table:pv-par}
\end{table*}

The carrier capture coefficient ($C_{n/p}$) is related to the capture cross section $\sigma$ and the thermal velocity $\nu$ via $C = \sigma\nu$. The \textit{ab-initio} computation of $\sigma$ is very expansive, thus we assume the same $\sigma$ for both charge carriers ($10^{-15} cm^{2}$) in both materials, a typical value used for deep defect states \cite{alkauskas2014first,shi2012ab}. The thermal velocity $\nu$ at particular temperature for electrons and holes is directly estimated from the carrier effective masses \cite{smets2016solar}. Moreover, the carrier densities ($\Delta n$) under illumination of light are kept fixed to $10^{15} cm^{-3}$ in both silicides, a typical value used in many PV materials such as CdTe \cite{yang2016non} and kesterite Cu$_{2}$ZnSnS$_{4}$ (CZTS) \cite{kim2019lone}. The $\varepsilon_{t}$ levels above the $\varepsilon_{v}$ for most stable defects in Mg$_{2}$Si and Ca$_{2}$Si are found to be $\sim$0.4 and 0.6 eV, respectively. Based on this information, in Table~\ref{table:pv-par}, the {\small{SRH}} lifetime ($\uptau$) is calculated to be $\sim$2 $\mu s$ for Mg$_{2}$Si, which is comparable to the lifetimes observed in various Cu-based solar materials \cite{dahliah2021high}. For Ca$_{2}$Si, the {\small{SRH}} lifetime is $\sim$11.3 $ms$, slightly higher than the 8.8 $ms$ lifetime for silicon \cite{park2018point}. Now, the {\small{SLME}} and $\eta_{{\small{TLC}}}$ at 300 K for both silicides, obtained using the BSE absorption spectrum, are presented in Fig.~\ref{fig:tlc}. The figure indicates that both values remain nearly the same up to a certain thickness. However, beyond this thickness, the {\small{TLC}} value drops below that of the {\small{SLME}} up to the studied thickness. This behavior, as understood from Eq.~\hyperref[eq:jnr]{\eqref{eq:jnr}}, can be attributed to significantly lower non-radiative recombination losses in thinner layers compared to thicker ones for both materials. Additionally, it is noted that the rate of decrease in {\small{TLC}} efficiency for Mg$_{2}$Si is greater than that for Ca$_{2}$Si, possibly due to a comparatively much lower $\uptau$ for Mg$_{2}$Si than for Ca$_{2}$Si. In Table~\ref{table:pv-par}, the highest {\small{SLME}} for Mg$_{2}$Si is $\sim$1.3\%, which is very low compared to the SQ-limit of $\sim$19.37\%. When considering non-radiative recombination, the highest $\eta_{{\small{TLC}}}$ is $\sim$1.2\%, showing no significant change compared to the {\small{SLME}}. One possible explanation for lower values is the large direct band gap ($\sim$2.45 eV), which results in a reduced contribution from the visible energy portion in the $\eta_{{\small{TLC}}}$ computation. This leads to a very low $J_{sc}$ value of $\sim$6.8 $mAcm^{-2}$, resulting in a minimal output power. In Table~\ref{table:pv-par}, the maximum $\eta_{{\small{TLC}}}$ of $\sim$28.5\% is lower than, as expected, the highest {\small{SLME}} of $\sim$31.2\% for Ca$_{2}$Si, and both values are below the SQ-limit of $\sim$31.25\%. The inclusion of non-radiative recombination losses reduces the highest efficiency value by only 2.7\% compared to the radiative limit ({\small{SLME}}). This {\small{TLC}} efficiency is comparable to that of silicon-based solar cells (26.7\%) \cite{park2018point}. Additionally, it is higher than the efficiencies of other materials, including CdTe (21\%), Cu(In,Ga)(S,Se)$_{2}$ (CIGS) at 21.7\%, halide perovskite (20.9\%) \cite{park2018point}, and various kesterite solar materials (21-24\%) \cite{kim2020upper}.

Based on these results, one can conclude that Ca$_{2}$Si can be effectively used to develop single-junction thin-film solar devices. In contrast, the significantly lower {\small{TLC}} efficiency of Mg$_{2}$Si makes it less suitable for single-junction applications. However, it may be effective as a bottom cell to absorb the lower-energy portion of the solar spectrum in multi-junction PV modules. It is worth noting that the obtained $\eta_{{\small{TLC}}}$ is associated with only one local single-defect level in the band gap. However, in reality, multiple local single-particle defect states can exist for a defect from various mid-gap transition levels \cite{alkauskas2016role}. The presence of these multiple trap levels significantly reduces $R_{{\small{SRH}}}$, as defects often become trapped in their highest or lowest charge states. Conversely, the inclusion of excited states can enhance the $R_{{\small{SRH}}}$ by several orders of magnitude, effectively balancing the former reduction \cite{alkauskas2016role}. Consequently, the overall calculated $\eta_{{\small{TLC}}}$ may not significantly change when accounting both factors in calculations. While accounting for recombination losses at the material interface \cite{smets2016solar} could influence the calculated $\eta_{{\small{TLC}}}$, such investigations are beyond the scope of this paper. Future studies could explore the impact of these multilevel and surface recombination loss mechanisms on {\small{TLC}} efficiency. 

\section{Conclusions}
In summary, we systematically investigate the electronic and optical properties of cheap, eco-friendly, and earth-abundant Mg$_{2}$Si and Ca$_{2}$Si materials using $first$ $principles$ based DFT and BSE methods, while also evaluating their solar performance factors. The calculated indirect and direct band-gaps are in the range of 0.25-0.6 eV and 0.57-0.96 eV for Mg$_{2}$Si and Ca$_{2}$Si, respectively. We also obtain the $m_{d}^{*}$ of electrons and holes at high-symmetric points for band extrema. In IPA, the $\epsilon_{1}(\omega)$ and $\epsilon_{2}(\omega)$ exhibit peak values of $\sim$50 (2.6 eV) and $\sim$61 (3.24 eV) for Mg$_{2}$Si, while for Ca$_{2}$Si, the values are $\sim$16.3 (1 eV) and $\sim$16.2 (3.4 eV), respectively. Inclusion of the excitonic effect in BSE solution raises the highest IPA peak of $\epsilon_{1}(\omega)$ ($\epsilon_{2}(\omega)$) to $\sim$59 (65) for Mg$_{2}$Si, and $\sim$17 (16.6) for Ca$_{2}$Si, with a shift towards lower energy. Both silicides demonstrate a large $\alpha(\omega)$ (order of 10$^{6}$ cm$^{-1}$) and $r(\omega)$ of less than 40\% in the active region of solar spectrum. The study illustrates the effectiveness of employing various theoretical methodology to accurately describe the properties of silicides. 

On the other hand, we have examined the performance of silicides by calculating efficiency based on (i) explicitly considering only radiative recombination using the {\small{SLME}} formalism and (ii) both radiative and non-radiative $e$-$h$ recombination using the {\small{TLC}} model. We find the interstitial Mg (Si) defect in Mg$_{2}$Si (Ca$_{2}$Si) to be more energetically favorable than other studied point defects. At 300 K, the highest {\small{SLME}} value is obtained to be $\sim$1.3 (31.2)\% for Mg$_{2}$Si (Ca$_{2}$Si), which is further reduced to be $\sim$1.2 (28.5)\% in the {\small{TLC}} calculation. The findings serve as useful guidelines for the experimental community in designing silicide-based solar devices and understanding their PV mechanisms.

	
	

\bibliography{paper}

\end{document}